# Halogen in Materials Design: Revealing the Nature of Hydrogen Bonding and Other Non-Covalent Interactions in the Polymorphic Transformations of Methylammonium Lead Tribromide Perovskite

Arpita Varadwaj,[a,b] Pradeep R. Varadwaj,*,†[a,b] Helder M. Marques,[c] Koichi Yamashita[a,b]

**ABSTRACT:** Methylammonium lead tribromide ($CH_3NH_3PbBr_3$) perovskite as a photovoltaic material has attracted a great deal of recent interest. Factors that are important in their application in optoelectronic devices include their fractional contribution of the composition of the materials as well as their microscopic arrangement that is responsible for the formation of well-defined macroscopic structures. $CH_3NH_3PbBr_3$ assumes different polymorphs (orthorhombic, tetragonal and cubic) depending on the evolution temperature of the bulk material. An understanding of the structure of these compounds will assist in rationalizing how halogen-centered non-covalent interactions play an important role in the rational design of these materials. Density functional theory (DFT) calculations have been performed on polymorphs of $CH_3NH_3PbBr_3$ to demonstrate that the H atoms on C of the methyl group in $CH_3NH_3^+$ entrapped within a $PbBr_6^{4-}$ perovskite cage are not electronically innocent, as is often contended. We show here that these H atoms are involved in attractive interactions with the surrounding bromides of corner-sharing $PbBr_6^{4-}$ octahedra of the $CH_3NH_3PbBr_3$ cage to form Br•••H(—C) hydrogen bonding interactions. This is analogous to the way the H atoms on N of the –$NH_3^+$ group in $CH_3NH_3^+$ form Br•••H(—N) hydrogen bonding interactions to stabilize the structure of $CH_3NH_3PbBr_3$. Both these hydrogen bonding interactions are shown to persist regardless of the nature of the three polymorphic forms of $CH_3NH_3PbBr_3$. These, together with the Br•••C(—N) carbon bonding, the Br•••N(—C) pnictogen bonding, and the Br•••Br lump-hole type intermolecular non-covalent interactions identified for the first time in this study, are shown to be collectively responsible for the eventual emergence of the orthorhombic geometry of the $CH_3NH_3PbBr_3$ system. These conclusions are arrived at from a systematic analysis of the results obtained from combined DFT, Quantum Theory of

* Corresponding Author's E-mail Address: pradeep@tcl.t.u-tokyo.ac.jp

[a] Dr A Varadwaj, Prof. P.R. Varadwaj, Prof. K. Yamashita, Department of Chemical System Engineering, School of Engineering, The University of Tokyo 7-3-1, Hongo, Bunkyo-ku, Tokyo, Japan 113-8656; and

[b] CREST-JST, 7 Gobancho, Chiyoda-ku, Tokyo, Japan 102-0076

[c]Prof. H.M. Marques,
Molecular Sciences Institute, School of Chemistry, University of the Witwatersrand, Johannesburg, 2050 South Africa

Atoms in Molecules (QTAIM), and Reduced Density Gradient Non-Covalent Interaction (RDG-NCI) calculations carried out on the three temperature-dependent polymorphic geometries of $CH_3NH_3PbBr_3$.

Keywords: Polymorphs of halide (bromide) perovskite; DFT calculations; Hydrogen bonding; Other non-covalent interactions; QTAIM and RDG-NCI analyses

**Graphical abstract**

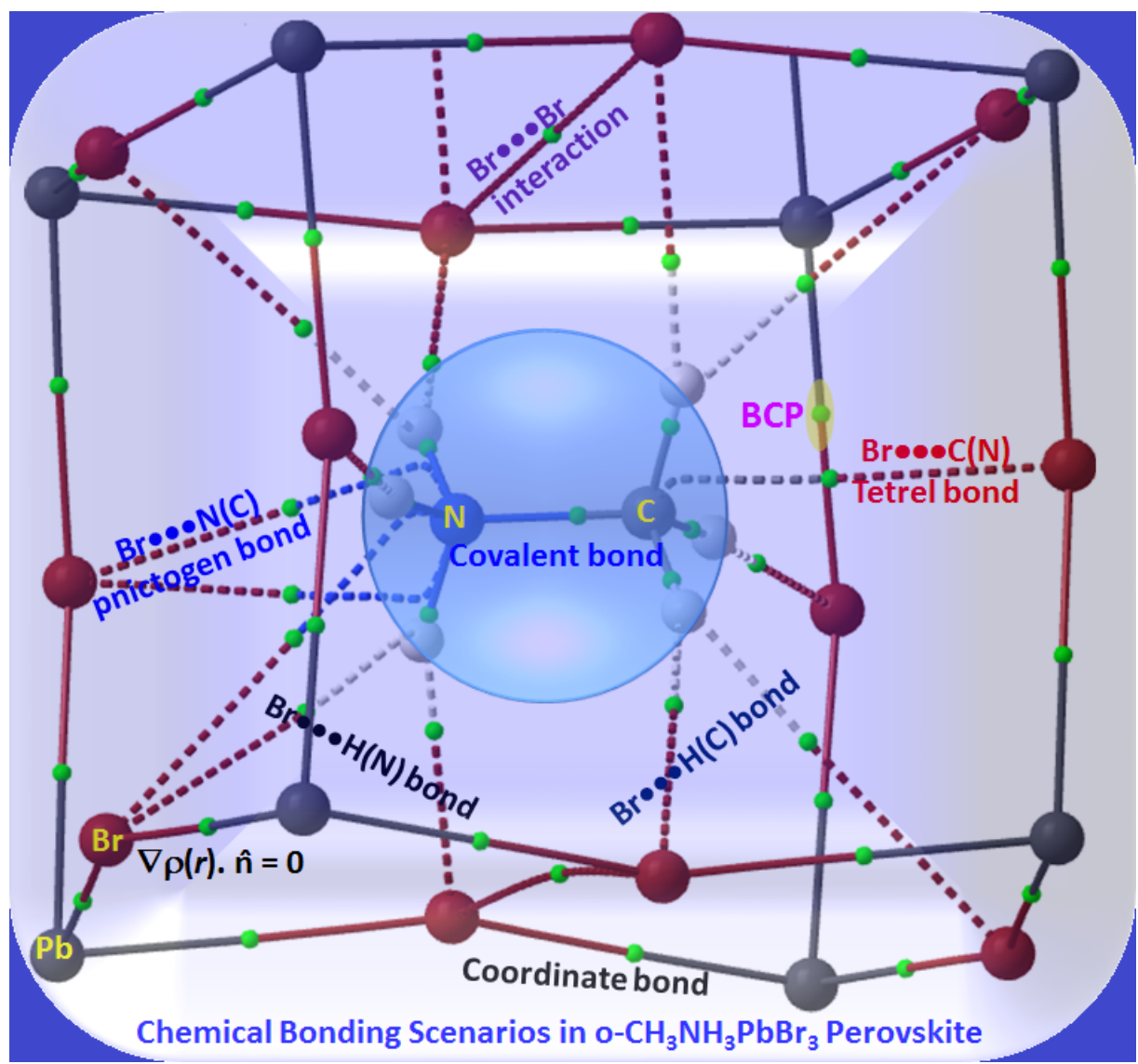

1.    Introduction

Halogen compounds in which halogen-assisted non-covalent interactions feature prominently have found diverse applications in crystal engineering [1], medicinal chemistry [2], supramolecular chemistry [3] materials [4-6] and drug design [7]. The halogens in molecules have the ability to donate electron density from electron rich lone-pair orbitals and to accept hydrogen atoms from other molecules to form hydrogen bonds, an important non-covalent interaction [8]. This can be usefully exploited in synthetic chemistry [3-6]. Hydrogen bonding does play an important role in close packing in crystals, and tunes the structural, electronic, optical and mechanical properties of diverse materials obtainable from, *inter alia*, π-conjugated molecular domains, and oligomeric and polymeric arrangements [5, 6].

Many criteria, characteristics and properties have been considered to identify and assign hydrogen bonding in molecules, molecular complexes, and solids, including, among others, geometric parameters, electron charge density distribution, infrared/Raman vibration frequency shifts, and NMR responses. In 2011, an IUPAC (International Union of Pure and Applied Chemistry) working group consolidated these and promulgated a set of recommendations for assigning hydrogen bonding interactions to further our understanding of these interactions in chemical systems [8].

In 2007, Clark *et al.* demonstrated that a covalently bound halogen X in a molecule can have a positive area on the outer extension of the R–X covalent bond [9]. Termed a σ-hole, it can attract a negative site and means that the charge density profile on the electrostatic surface of a covalently bound halogen in a molecule is anisotropic. Consequently, the lateral sites of a covalently bound halogen derivative can serve as a hydrogen bond/halogen bond acceptor while the axial site can serve as a halogen bond donor. Since then, there have been almost countless research articles published in the area of halogen bonding, demonstrating its profound importance in materials and drug design.

In 2013, IUPAC recommended another set of criteria to correctly assign and analyze halogen bonding in chemical systems [10]. The same idea and characteristic features have been extended to the chemical bonding interactions formed by the elements of Groups 14, 15 and 16 of the periodic table, leading to definitions of tetrel bonding [11], pnictogen bonding [12, 13], and chalcogen bonding [12, 13], respectively, even though these had been crystallographically identified (but unnamed [13]) over the decades.

Just as in DNA base pairs [14], (halogen-assisted) hydrogen bonding interactions are very important in organic-inorganic hybrid halide-based perovskites [15-27]. The geometrical shape, stability and functionality of these functional materials have no physical basis without these non-covalent interactions.

Methylammonium lead tribromide ($CH_3NH_3PbBr_3$ or $MAPbBr_3$) is the second largest member of the $BMY_3$ (B = $CH_3NH_3^+$, M = Pb/Sn/Ge, and Y = Cl, Br, I) perovskite family [28, 29]. The excellent coupling between a phonon, the $CH_3NH_3^+$ organic cation, and the $MY_3^-$ inorganic lattice in this system enables this material to serve as a light absorber. Considerable research on these compounds is on-going to address many fundamental and technological issues, but the details of the chemical bonding topology in these compounds are often misunderstood, misinterpreted or underestimated. The IUPAC recommended criteria [8], as well as similar criteria well-established by other research groups [9-13], for hydrogen bonding (and other non-covalent interactions) are not always taken into account when examining their presence in these systems. It is our view that these interactions merit careful study because the structural stability and functionality of perovskite materials (whether organic-inorganic, all-inorganic or metal-organic) hinge on these non-covalent and hydrogen bonding interactions between the guest and the perovskite host, the latter composed of dative coordinate bonding interactions. For example, it has been realized that the hydrogen bonding interactions between the H atoms of the organic cation and the bromides of the host lattice provide the structural stability of the $MAPbBr_3$ system [24, 30]. Some of us have recently established that the gas phase binding energies calculated for the $MAPbY_3$ systems correlate well with the experimentally determined optical bandgaps for these thin-film materials, and serve as a tool for the fundamental understanding of the chemistry and design strategy of these materials [24].

Yoon *et al.* have shown that the stability of the complexation between $Pb^{2+}$ and $Br^-$ is nearly 7 times greater than that for complexation between $Pb^{2+}$ and $I^-$, thus making $Br^-$ a dominant binding species in mixed halide systems [31]. It would be interesting to determine what causes the Br-substituted perovskite material to be more photostable compared to methylammonium lead triiodide ($CH_3NH_3PbI_3$). One way to obtain insight into this specific question is to explore and understand the detailed chemical bonding environment in $CH_3NH_3PbBr_3$ and compare and contrast it with that of $CH_3NH_3PbI_3$.

We stress that the importance of non-covalent interactions in the $MAPbBr_3$ system has recently been investigated by others [22]. The results of combined experimental work, DFT and QTAIM calculations were used to provide insight into the nature of the evolution of hydrogen bonding in the three polymorphs mentioned above. The QTAIM analysis of the study suggested the presence of two types of hydrogen bonding interactions in orthorhombic

MAPbBr$_3$ (o-MAPbBr$_3$), referred to as H$_N$···Br and H$_C$···Br, where H$_N$ and H$_C$ refer to hydrogen atoms of the amino and methyl fragments, respectively, of the CH$_3$NH$_3^+$ organic cation. It was suggested that the H$_C$···Br bond becomes significant only in the low-temperature orthorhombic polymorph. For the tetragonal and pseudocubic geometries of the system only H$_N$···Br hydrogen bonds were identified and H$_C$···Br hydrogen bonds (and other non-covalent interactions) were found to be absent [22].

This conclusion [22], reached for o-CH$_3$NH$_3$PbBr$_3$ is in sharp contrast with what was concluded previously for the o-CH$_3$NH$_3$PbI$_3$ system reported by others [16-19, 21, 25, 26], the heaviest member of the BMY$_3$ perovskite series [24]. For instance, Lee *et al.* [16] considered only the H$_N$···I hydrogen bonds formed by the ammonium fragment of the organic cation with the coordinated iodides. They attempted to link these hydrogen bonds with the tilting of the PbI$_6^{4-}$ octahedra that controls the geometrical and photovoltaic properties of o-CH$_3$NH$_3$PbI$_3$, but the importance of H$_C$···I hydrogen bonds was completely neglected. This has prompted us to thoroughly examine the o-CH$_3$NH$_3$PbBr$_3$ system to delineate the details of the bonding interactions that are responsible for this polymorph.

In this study, we have performed electronic structure calculations for the orthorhombic (*Pnma*), tetragonal (*I4/mcm*) and pseudo-cubic ($Pm\overline{3}m$) phases of MAPbBr$_3$ with periodic Density Functional Theory (DFT) using the Vienna *Ab initio* Simulation Package (VASP) [32-34]. The equilibrium geometries obtained from these calculations were compared with those available experimentally from neutron diffraction measurements [15].

The principal focus of this study was to identify and characterize non-covalent bonding interactions in MAPbBr$_3$ and to determine whether there is any connection between octahedral tilting and hydrogen bonding in the low temperature phase as has been suggested for MAPbI$_3$ [16, 17]. We have also investigated the nature of potential intermolecular interactions between the organic and inorganic moieties that are solely responsible for the stability of the high temperature polymorphic geometries the system. Towards this end, Quantum Theory of Atoms in Molecules (QTAIM) [35-40] and Reduced Density Gradient Non-Covalent Interaction (RDG-NCI)) [41] calculations were performed on each of the equilibrium geometries of the three phases of the MAPbBr$_3$ system obtained from DFT calculations. Analogous studies on the CH$_3$NH$_3$PbI$_3$ system are currently underway, and will be reported elsewhere.

One of the aims of this study is to show that just using distance cut-offs to explore hydrogen bonding interactions in CH$_3$NH$_3$PbBr$_3$ can be misleading and can lead to an incomplete understanding of the system being studied [23]. Second, we show that the hydrogen bonding interactions H$_C$···Br and H$_N$···Br are ubiquitous in all the polymorphic geometries of the MAPbBr$_3$ system. Third, and using other examples from the literature [42-46], we provide a description of "hydrogen bonds" that can be formed between a hydrogen involved in a low-polarity bond such as a C–H bond, and a negative site. Fourth, we elucidate that in addition to these two types of hydrogen bonds there are several other non-covalent interactions between MA and PbBr$_6^{4-}$ that are also responsible for the tilting of the PbBr$_6^{4-}$ octahedra in o-MAPbBr$_3$, as well as the structural stability of pseudocubic MAPbBr$_3$ (c-MAPbBr$_3$) that has the organic cation orientation distorted along the [100] direction. We have done this by utilizing a number of the topological signatures of the charge density that emerge from QTAIM [35-40] and RDG-NCI [41], and a few IUPAC-recommended criteria for hydrogen bonding [8] and halogen bonding [9, 10]; we have also used established criteria to identify other non-covalent interactions including carbon bonding [11], chalcogen- and pnictogen-bonding [12, 13] and lump-hole bonding [47] to arrive at a comprehensive description of the non-covalent interactions in the MAPbBr$_3$ system.

## 2. Computational Methods

We have used the standard projector augmented-wave (PAW) potential method [48, 49] and the Perdew–Burke–Ernzerhof exchange-correlated functional [50, 51] to obtain the equilibrium bulk and supercell geometries of the orthorhombic (*Pnma*), tetragonal (*I4/mcm*) and pseudo-cubic ($Pm\bar{3}m$) phases of the CH$_3$NH$_3$PbBr$_3$ system. These were obtained by sampling the Brillouin zone for respective geometries with 10 × 8 × 10, 8 × 8 × 10, and 12 × 12 × 12 *k*-point meshes centered at $\Gamma$, respectively. The plane-wave cut-off energy and convergence criterion for energy were set to 520 eV and 10$^{-5}$ eV/Å, respectively. These calculations were performed using the VASP code [32-34]. A similar computational procedure and settings were previously used by others [22] to study the CH$_3$NH$_3$PbBr$_3$ system, but the Brillouin zone was sampled with 6 × 6 × 6 (pseudo-cubic) and 4 × 4 × 2 (orthorhombic and tetragonal) *k*-point meshes centered at $\Gamma$.

Gaussian 09 [52] was used to generate wavefunctions from the geometries of the periodic bulk and supercells of the perovskites for QTAIM charge density analysis [40]. To our knowledge, QTAIM, which relies on the zero-flux boundary condition [35-40], is one of the current leading theoretical approaches used to study the nature and strength of hydrogen bonding and other non-covalent interactions. Two of its charge density-based topological descriptors, *viz.*, the presence of a bond path (bp), and the presence of the (3,–1) bond critical point (bcp) between interacting atomic basins have proved very useful in inferring the presence of a chemical bonding in chemical systems [35-39, 53-60]. Two of its other signatures, the sign of the Laplacian of the charge density ($\nabla^2\rho_b$) and the sign of the total energy density ($H_b$) at a bcp, can be used (and have been recommended) to determine whether the interaction between two atomic basins in a given pair is ionic, covalent, or a mixture of the two [59-65]. We have suggested in a recent study that these, together with other indicators for non-covalent interactions, can be used in the unambiguous characterization of hydrogen bonding in chemical systems [61]. These QTAIM signatures have been employed by others to explore chemical bonding in MAPbBr$_3$ [22]. They used a range of values that have been previously suggested for $\rho_b$ and $\nabla^2\rho_b$ (0.002 < $\rho_b$ < 0.034 au; 0.024 < $\nabla^2\rho_b$ < 0.139 au) to identify and characterize O···H(–C) hydrogen bonding in molecular complexes [66]; however, as we will show below, these may indeed be a useful *guide*, but should not be treated as a *prerequisite* for the identification of hydrogen bonding in chemical systems since non-covalent interactions, in general, do not have strict boundaries [42, 43, 46, 58]. We will also demonstrate that because of its stringent criteria QTAIM sometime fails to identify weakly bound and van der Waals interactions in chemical systems [67-71]; this is the reason we employed the RDG-NCI approach to investigate the details of the bonding topologies involved.

3. Results and Discussion

There is emerging evidence from a variety of experimental techniques that MAPbBr$_3$, as well as MAPbI$_3$, has a soft lattice with rich dynamics including rotational motion of the MA cation, although only at or above room temperature. There is some understanding that the dynamic tilting modes of PbI$_6^{4-}$ octahedra are coupled to the movement and dynamics of the CH$_3$NH$_3^+$ cation, a coupling that originates from the hydrogen bonding interactions between the cationic guest and anionic host lattice [30]. Even without taking molecular dynamics into account, it is not difficult to provide insight into what relevance the pattern of the non-covalent interactions that emerge have on the structures of the perovskite examined. The analysis of hydrogen bonding in these systems has been carried out previously by exploring the static equilibrium and/or the experimental geometries of the various phases, regardless of whether these have been determined crystallographically (by X-ray or neutron diffraction) [16-21] or by computational methods using periodic DFT calculations [16-18, 22, 23, 47], and the nature of specific patterns of hydrogen bonding in the geometries of various phases have been proposed. Moreover, since the organic cation in o-MAPbBr$_3$ is experimentally observed to be in a fixed orientation in the low temperature phase, a molecular dynamics simulation is deemed unnecessary to provide insight into the nature of the non-covalent interactions involved. Both MAPbI$_3$ and MAPbBr$_3$ have previously been explored by molecular dynamics simulations [72-75], yet insight into the underlying bonding topology is, in our view, and those of others [30], still greatly lacking. We believe that the results presented below on the equilibrium static geometries significantly enhances our current understanding of the bonding scenarios in the MAPbBr$_3$ system.

The widely used periodic DFT-PBE functional is suitable for modeling solid state materials and is adequate to describe the equilibrium geometry, intermolecular interactions, and bandgaps in halide perovskites [76]. Since dispersion is a major issue in such systems because of the involvement of several heavy atoms (i.e., Br), we have considered the effect of van der Waals interactions by introducing a non-local van der Waals correction into the DFT calculations such as the DFT-D3 method with Becke-Johnson damping [77, 78]. When using this procedure, we observed that the bond distances and bond angles associated with the hydrogen bonds are only slightly decreased, but this did not affect the general conclusions arrived at in the following section. Using different functionals (*viz*. PBEsol and PBEsol-D3 implemented in VASP [32-34]) led to slightly different inter-atomic distances for the non-covalent interactions we described below, but again the overall conclusions did not change.

**Orthorhombic MAPbBr$_3$**

Figure 1 shows the PBE relaxed geometry, lattice constants and unit cell volume of o-MAPbBr$_3$. The optimized lattice constants are comparable with those reported experimentally, *viz.*, $a$ = 7.976, $b$ = 11.841 and $c$ = 8.565 Å, belonging to the nonpolar *Pnma* space group.[79]   The MA cation in the relaxed geometry is in a staggered configuration, with its center of mass lying near the center of the inorganic perovskite cage, in agreement with experiment [15, 19-21, 27]. The bandgap, which is the difference between the energy levels of the valence band maximum and conduction band minimum, is calculated to be 2.16 eV, in reasonable agreement with the experimental bandgap of 2.33 eV [80, 81].

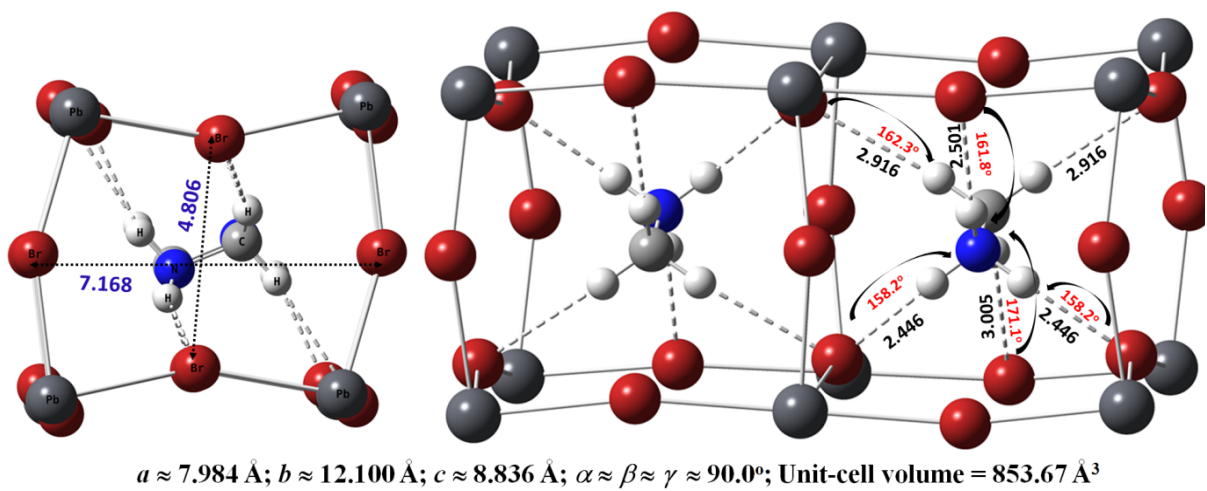

$a \approx 7.984$ Å; $b \approx 12.100$ Å; $c \approx 8.836$ Å; $\alpha \approx \beta \approx \gamma \approx 90.0°$; Unit-cell volume = 853.67 Å$^3$

**Figure 1.** Two views of the ball and stick model of the PBE relaxed geometry of o-MAPbBr$_3$, showing some tentative intermolecular bond-distances and bond-angles of approach for the formation of intermolecular contacts (in Å and deg, respectively). Optimized lattice constants and volume are shown. Figure S2 gives the polyhedral views. Atom labeling is shown in the ball and stick model on the left.

The three H atoms of the –NH$_3^+$ group in MA are attracted toward the surrounding bromides of the perovskite cage in o-MAPbBr$_3$, forming at least three Br···H(–N) intermolecular contacts. This is evidence of the occurrence of intermolecular contacts; two of these contacts are short and equivalent, with $r$(Br···H(–N)) = 2.446 Å while the remaining one is longer, with $r$(Br···H(–N)) = 2.501 Å. These are hydrogen bonds since each of these three intermolecular distances is significantly smaller than the sum of the van der Waals radii of the Br and H atoms, 3.06 Å (Br = 1.86 Å and H = 1.20 Å [82]); this is a generally-accepted [44] concept for hydrogen bonding and a IUPAC-recommended [10] characteristic for identifying halogen bonds. This result indicates that there is appreciable mutual penetration between the non-covalently bound Br and H atoms involved in hydrogen bond formation in this system [44, 45, 83, 84]. These hydrogen bonds between the inorganic host and organic guest species in o-CH$_3$NH$_3$PbI$_3$ may be incipient states of the proton transfer processes that accompany ring opening and closure, making the entire structure compact and stable. While this view is in line with that of Bürgi and Dunitz for hydrogen bonds [85], Desiraju and Steiner are of the opinion that hydrogen bond in general can be regarded as the incipient state of a proton transfer process, irrespective of whether it is strong or weak, and that it is only in the case of strong hydrogen bonds that such proton transfer processes occur at significant rates [44].

The H atoms of the –CH$_3$ group of the MA cation also form intermolecular contacts with the Br atoms of the perovskite cage in o-CH$_3$NH$_3$PbI$_3$. This is not unexpected given the staggered nature of the MA cation, Figure 1. As expected, there are two Br···H(–C) H-bonded contacts that are equivalent and the other that is slightly longer, $r$(Br···H(–C)) values 2.916 and 3.005 Å, respectively. Each of these contact distances is slightly smaller than the sum of the van der Waals radii of the H and Br atoms, 3.060 Å, in line with the IUPAC recommendation [10]. This signifies that the hydrogen bonds formed between the H atoms of the –$CH_3$ group and the Br-atoms of the perovskite cage are not very weak.

A study by Ryan *et al.* [86] gives an example of an analogous non-covalent interaction motif. They reported the crystal structure of $[C_6H_5NH(CH_3)_2]_2Te_2I_{10}$. It consists of the *N,N*-dimethylanilinium cation and a hitherto unreported tellurium iodide anion $Te_2I_{10}^{2-}$. The $Te_2I_{10}^{2-}$ dianion is accompanied with two edge-sharing $TeI_6^{2-}$ octahedra. It builds up a three-dimensional Te(IV)-I open framework through extensive interconnecting I···I non-covalent contacts. The intermolecular distances for these contacts ranged between 3.66 and 3.80 Å, shorter than twice the van der Waals radius of the iodine atom (4.04 Å). It was suggested that these contacts may potentially promote charge carrier migration throughout the Te—I network [86].

Neither the three Br···H(–N) nor the three Br···H(–C) bond distances are equivalent in $CH_3NH_3PbBr_3$, Figure 1. This accords with what was found in the crystal structure of o-$CH_3ND_3PbBr_3$ [15]: two of the Br···D(–N) deuterium bond distances formed by the ammonium D atoms were 2.525 Å, while the other was 2.500 Å. Similarly, two of the Br···H(–C) hydrogen bond distances formed by the methyl H atoms were. 2.916 Å and the other one was 2.988 Å. This is also in agreement with the result of Lee *et al.* [16] who reported non-equivalent I···H(–N) hydrogen bonds for o-MAPbI$_3$; they then connected these with the tilting of $PbI_6^{4-}$ corner sharing octahedra, described by the Glazer notation $a^-b^+a^-$ [87], where the superscripts + and – represent in-phase and anti-phase tilts, respectively, along *a* and *b*, the two mutually perpendicular principal crystallographic axes. The result of Lee *et al.* [16] contrasts with that of Yin *et al.* [22] in that the former completely neglects the importance of the I···H(–C) hydrogen bonds in o-$CH_3NH_3PbI_3$, whereas the latter recognizes the significance of the analogous Br···H(–C) hydrogen bonds in the geometrical stability of o-$CH_3ND_3PbBr_3$.

A lengthening in the C—N bond distance in MA is observed upon going from the room temperature tetragonal to the low-temperature orthorhombic geometry of MAPbBr$_3$. This is intuitively correct since the MA cation has a higher propensity for hydrogen bond formation with the Br atoms of the inorganic perovskite cage in o-MAPbBr$_3$ than in t-MAPbBr$_3$. There is experimental precedence for this in MAPbI$_3$ [19]; there are two equivalent and one short

I⋯H(–N) hydrogen bonds in o-MAPbI$_3$ (2.808(9) and 2.613(7) Å, respectively), whereas for the tetragonal phase these are at 3.18(1) Å and 3.15(2) Å. The longer hydrogen bond distances in the tetragonal phase are indicative that the intermolecular coupling between the inorganic and organic moieties in the MAPbBr$_3$ geometry of this phase is weaker than in the low temperature orthorhombic phase. This may explain why the overall strengths of the Br⋯H(–N) and Br⋯H(–C) hydrogen bonding interactions in MAPbBr$_3$ are stronger in the orthorhombic than in the tetragonal and cubic phases (*vide infra*).

The Br⋯H(–N) hydrogen bonds, on the other hand, are stronger than the Br⋯H(–C) hydrogen bonds in o-MAPbBr$_3$ if inter-atomic distance is used as the sole criterion since the former are shorter than the latter. However, from the angle of approach for hydrogen bond formation (Figure 1), it can be concluded that the Br⋯H(–C) hydrogen bonds are more directional than the Br⋯H(–N) hydrogen bonds. The same can also be inferred from the crystal structure of CH$_3$ND$_3$PbBr$_3$ (Figure S1 of the ESI) [15]. It is generally believed that the more linear (directional) non-covalent interactions are, the stronger they are. Thus, it would be judicious to conclude that long and strongly directional Br⋯H(–C) hydrogen bonds may be competitive with short and weakly directional Br⋯H(–N) hydrogen bonds in o-MAPbBr$_3$. This means that the –NH$_3$ and –CH$_3$ groups of the CH$_3$NH$_3^+$ cation are strongly competitive with each other in forming hydrogen bonded interactions with the cage bromides in o-MAPbBr$_3$. This competition between the two groups of MA in forming two types of hydrogen bonds actually pulls the C–N bond symmetrically in opposite directions, causing its marginal elongation in o-MAPbBr$_3$ compared to that found in t-MAPbBr$_3$, thus explaining what Yin *et al.*[22] found puzzling. Unless this view is correct, then it is difficult to understand why, unlike in the two room temperature phases, the organic cation should reside almost at the center of the perovskite cage in o-CH$_3$ND$_3$PbBr$_3$; one would expect it to be off-center because of the high propensity of the H atoms of the –NH$_3^+$ fragment to engage in non-covalent interactions.

Figure 2 shows the QTAIM molecular graph for o-MAPbBr$_3$. As expected, it has captured all the coordinate bonding interactions between Pb and Br atomic basins. Each Pb–Br coordinate bond is characterized by a single bcp and a pair of gradient paths. By definition, each of these two gradient paths originates and terminates at the (3,–1) bcp, and hence a pair of these produces the bp that links the nuclei of the bonded Pb and Br atom basins. The presence of bcp's and bp's between the atomic basins is a general topological property of molecular charge distributions [38].

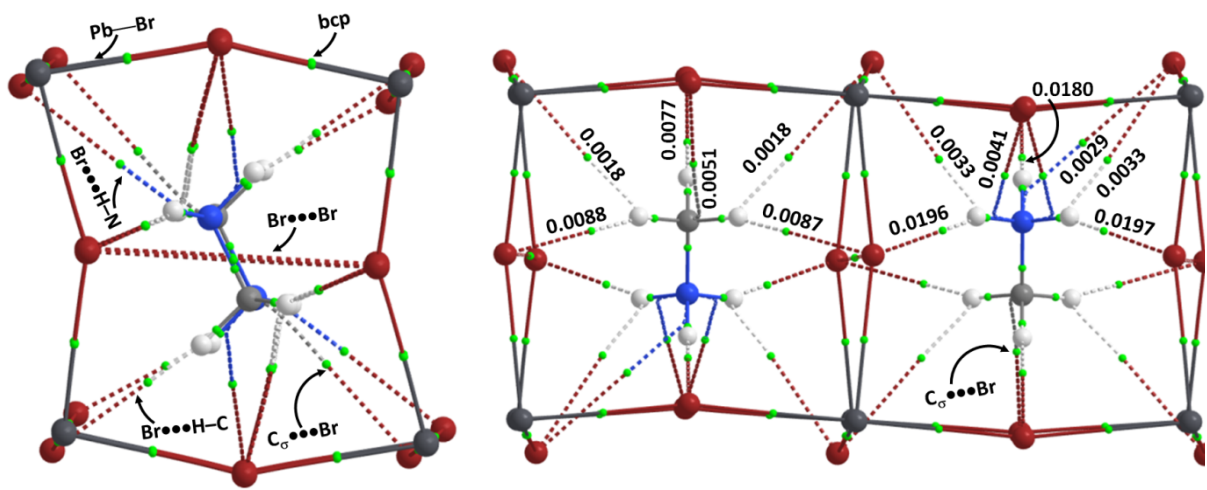

**Figure 2.** Two different views of the PBE QTAIM molecular graph of the bulk relaxed geometry of o-MAPbBr$_3$ (48 atoms), illustrating bcp's (tiny spheres in green) and bond paths (solid and dotted lines in atom color) between various atomic basins interacting with each other. Charge density values at some selected bcp's are shown. Atoms are shown as large and medium spheres. Lead: dark-grey, bromine: dark-red, nitrogen: blue, carbon: grey, and hydrogen: white grey. Because of symmetry, the $\rho_b$ values at Br⋯H(–N) and Br⋯H(–C) bcps are shown in two different distorted cubes.

The Pb–Br charge densities at the bcp's are calculated to be as large as 0.027 and 0.029 au, with the Laplacian of the charge densities $\nabla^2\rho_b > 0$. The potential energy density $V_b$, compared to the gradient kinetic energy density $G_b$, is potentially stabilizing at each of these bcp's, so that the total energy density $H_b$ is negative ($H_b = V_b + G_b < 0$). These attributes ($\nabla^2\rho_b > 0$ and $H_b < 0$) signify that the dative coordinate bonding interactions between Pb and Br have mixed ionic and covalent character [59-65]. There is no indication of multiple bonding in any of the coordination bonds observed between the Pb and Br atomic basins in o-MAPbBr$_3$ since the ellipiticity[35-39, 53] (0.002–0.010) and electron delocalization indices [88] (0.4–0.5) for the Pb–Br bonds are indicative of single bonds.

The QTAIM results suggest three very strong Br⋯H(–N) hydrogen bonding interactions between the –NH$_3^+$ group and Br atoms in CH$_3$NH$_3$PbBr$_3$, in agreement with those inferred from radii-based rationalizations (*vide supra*). The strength of these interactions can be quantified by the values of $\rho_b$ at the bcp's, which are 0.0197 and 0.0180 au for the two equivalent and one longer hydrogen bonds, respectively (see Figure 1 for bond distances and Figure 2 for $\rho_b$ values). The sign and magnitude of $\nabla^2\rho_b$ at the corresponding bcps are +0.038 and +0.033 au, respectively, underlining the electrostatic nature of these interactions. The total energy density at the corresponding bcp's are all found to be negative ($H_b = -0.0007$ au). These latter two signatures suggest that the Br⋯H(–N) hydrogen bonds have mixed character, *viz.*, they are significantly ionic with some covalent component [60-62].

In addition to these three primary Br⋯H(–N) hydrogen bonding interactions the QTAIM results also reveal several other secondary interactions between the H atoms of the –NH$_3^+$ group and the Br atoms of the perovskite cage in o-CH$_3$NH$_3$PbBr$_3$. Appearance of these interactions is not very surprising given the relatively small host/guest volume ratio of the overall negative perovskite cage interior and the cationic guest. Had the cage been larger (or the guest smaller), it might have been possible for the positively charged organic cation to attract a specific portion of the cage, as has been observed in many endohedral fullerene systems [89]. As it is, the space inside the PbBr$_6^{4-}$ cage is just adequate to encapsulate the guest cation, and this is probably the reason the B-site species is experimentally observed to be localized almost at the center of the PbBr$_6^{4-}$ cage in o-MAPbBr$_3$ [15, 22]. This close size match is probably one of the reasons for the formability of the orthorhombic perovskite stoichiometry; this can also be explained by Goldschmidt tolerance factor [90, 91]. Nevertheless, and as depicted in Figure 2, each H atom in –NH$_3^+$ is involved in the formation of at least two Br⋯H(–N) hydrogen bonds, one stronger and one weaker, with the latter ones characterized by smaller $\rho_b$ values at their bcp's.

A similar charge density topology is revealed for the bonding interactions of the –CH$_3$ group of MA in o-MAPbBr$_3$. Each of the two equivalent Br⋯H(–C) contacts identified through distance-based criterion is split into a bifurcated topology of bonding. The $\rho_b$ at the bcp's of these interactions are 0.0018 and 0.0088 au (Figure 2).

The carbon atom along the outer portion of the N–C σ-bond in MA is involved in an attractive interaction with the nearest bromide atom of the perovskite cage, forming a Br⋯C$_\sigma$–N type σ-hole centered carbon bonding interaction. This is identified for the first time in this study; the importance of this kind of bonding interaction in materials design has been at the center of a number of recent discussions [11, 13, 92-94]. Although this interaction is somewhat weaker in the calculated geometry of o-MAPbBr$_3$, it is reasonably significant in the experimental geometry of o-CH$_3$ND$_3$PbBr$_3$ (Figure S3). This can be understood by looking at the $\rho_b$ values summarized in Figure 2 and in Figs. S3-S4.

As shown in Figure 2, there are two bond paths that are developed between the nitrogen outer surface and the coordinated bromides in o-MAPbBr$_3$. These can be regarded as Br⋯N–C type σ-hole centered pnictogen bonding [12, 13, 93, 95]. These contacts are formed because the outer portion of the 0.001 au isoelectron density mapped electrostatic surface of nitrogen/carbon along the extension of the C–N/N–C σ-bond in MA is found to be positive that attracts the negative site to form a non-covalent interaction.

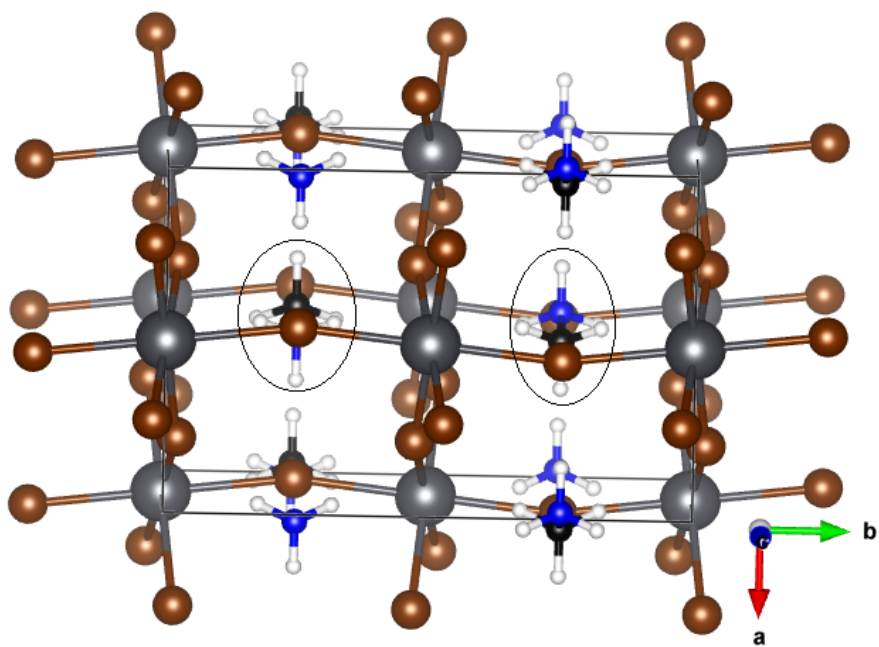

**Figure 3.** Alignment of the organic cation MA in the PBE relaxed geometry of bulk o-MAPbBr$_3$. This alignment of MA in the computed relaxed geometry is homologous to that found in the experimental geometry of o-CH$_3$ND$_3$PbBr$_3$ (*cf.* Figure S1), which is responsible for the development of Br···C–N and Br···N–C carbon- and pnictogen-bonding interactions, respectively.

The presence of both the Br···N–C and Br···C–N bonding arrangements maximizes the non-covalent interaction between the organic and inorganic moieties, causing the ordered alignment of the organic species between the bromine atoms of the PbBr$_6^{4-}$ octahedra in o-MAPbBr$_3$; this is shown by circles marked in Figure 3. Figure 2 illustrates the charge density topological details of these non-covalent interactions.

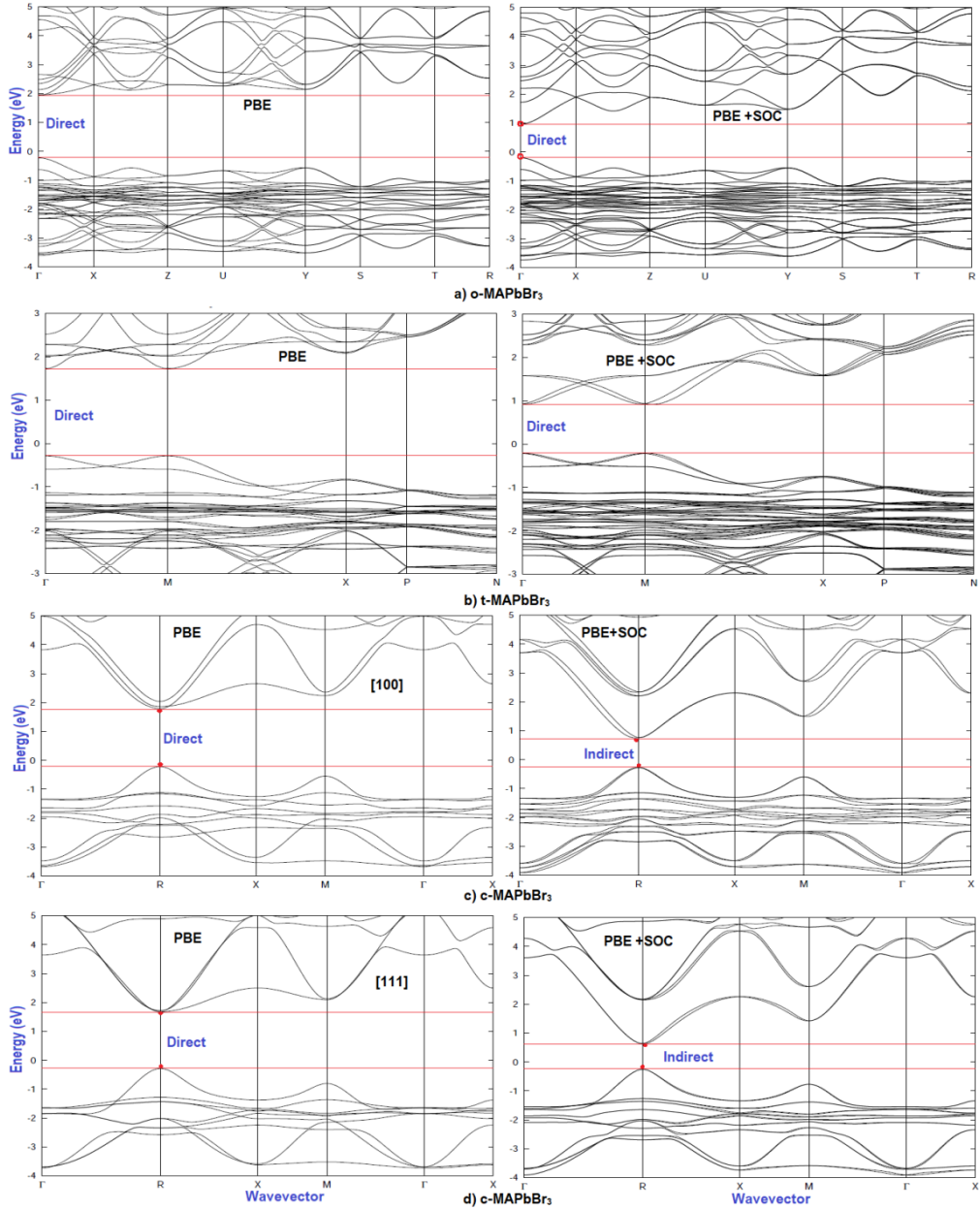

**Figure 4**: A comparison of the PBE level (without (left) and with-SOC (right)) electronic band structures for the MAPbBr$_3$ perovskites in the a) orthorhombic, b) tetragonal and c)-d) pseudo-cubic phases. The bandgap is shown to be direct at high symmetry $\Gamma$-, $\Gamma$-, and *R*-points for the three polymorphs without SOC, respectively. Inclusion of the effect of SOC has caused a spin-splitting of the conduction band for t-MAPbBr$_3$, c-MAPbBr$_3$ [110] and c-MAPbBr$_3$ [111], respectively, leading to the direct-to-indirect bandgap transition. This is the combined consequence of the non-centrosymmetric nature of the local geometries of these polymorphs driven by hydrogen bonding and the relativistic Rashba-Dresselhaus splitting of the lower conduction band.

In addition to the strong network of σ-hole interactions observed between MA and PbBr$_6^{4-}$ octahedra, other types of non-covalent interactions appear to exist between the inorganic and organic frameworks of o-MAPbBr$_3$. For instance, and as shown in Figure 2 (left), the axial and equatorial Br⋯Br distances of separation in o-MAPbBr$_3$ are calculated to be 4.806 and 7.188 Å, respectively. While the latter distance is very large, QTAIM predicts the former to have the characteristics of an attractive interaction. There are three such Br⋯Br attractive interactions feasible in bulk o-MAPbBr$_3$, and are dispersion driven. Previous studies show that non-covalent interactions of this type can be described by the lump-hole theory [47], as well as by London dispersion [96-100].

We have carried out similar calculations on o-MAPbBr$_3$ with PBE-D3, PBEsol, and PBEsol-D3 as implemented in VASP [32-34] to see how the use of these functionals affects the intermolecular distances and other properties. In all cases, we found that these distances decreased compared to those evaluated with PBE. In addition, the lattice constants were slightly affected, and hence so was the bandgap of the material. For instance, the respective two equivalent short and one longer $r$(Br⋯H(–N)) distances were 2.4072 and 2.4051 Å with PBE-D3, 2.3880 and 2.3977 Å with PBEsol, and 2.3473 and 2.3469 Å with PBEsol-D3. Similarly, the two equivalent long and one shorter $r$(Br⋯H(–N)) distances were 2.8181 and 2.8997 Å with PBE-D3, 2.8349 and 2.8960 Å with PBEsol, and 2.7528 and 2.8188 Å with PBEsol-D3. The bandgaps were 2.05, 1.79 and 1.91 eV with PBE-D3, PBESol and PBESol-D3, respectively. This is not unexpected given that the bandgap for halide perovskites is generally affected by small changes in the electronic structure, unit-cell volume and lattice constants. Despite the different functionals used, the conclusions reached for the nature of the intermolecular interactions in o-MAPbBr$_3$ (as well as those for the other room temperature polymorphs discussed below) and characterized with QTAIM remain the same.

Indirect to direct (or direct to indirect) bandgap transition in Pb-based perovskite solar cell materials[101-105] is a consequence of the Rashba-Dresselhaus effect [106], irrespective of whether the material in question is MAPbBr$_3$ or MAPbI$_3$. The effect is related to the splitting of conduction and valence band energy levels in momentum space [106]. It generally shows up only when the effect of spin-orbit coupling (SOC) is taken into account since it plays an important role in chemical systems that involve heavy atoms, as in MAPbBr$_3$. Whalley *et al.* have recently noted that SOC is not expected to have a large impact on the structural properties of the Pb-based compounds as it is the (empty) conduction band that is mainly affected [105].

For MAPbX$_3$ (X = Cl, Br, I) perovskites the direct-to-indirect bandgap transition occurs only for the pseudo-cubic geometries (and not in o-MAPbX$_3$) [107], and this shows up only when the geometries deviate locally from their ideal $Pm\bar{3}m$ symmetry. Because of the local geometrical distortions there is a breaking of the center of inversion symmetry. The compound then adopts a symmetry such as *P1*. Whalley *et al.* have suggested that this is necessary since it is a prerequisite for generating a local electric field . Thus as mentioned above, the Rashba-Dresselhaus effect (band splitting) should not occur in the orthorhombic crystal of MAPbBr$_3$ since its structure is highly symmetric. This means that whereas the H$_C$···I and other non-covalent interactions bond explain the state of the MA cations and concomitant tilting of the PbBr$_6^{4-}$ octahedra in o-MAPbBr$_3$ with the subsequent emergence of the band structures, there should not be any such direct-to-indirect bandgap transition in o-MAPbBr$_3$ as previously suggested.[22] Figure 4 clarifies this, where a comparison between the electronic band structures of o-, t- and c-MAPbBr$_3$ is provided.

In essence, it is logical to conclude that the chemical bonding topology in o-MAPbBr$_3$ is considerably more complex than in, for example, (NH$_3$)$_2$ [108] and (OH$_2$)$_2$ dimers [109]. Electron density theories such as QTAIM[35] and others (for example, RDG-NCI [41] and DORI [110]) are alternative theoretical approaches that can be utilized to faithfully assist uncovering the bonding topology, and importantly, this is when most of the other approaches fail. In the worst case, if we were arbitrarily to impose a cut-off for electron density at bcp's below 0.0060 au to neglect the presence of all manner of weakly bound non-covalent interactions between MA and PbBr$_6^{4-}$, then QTAIM would suggest at least six intermolecular hydrogen bonded interactions are formed between the organic and inorganic moieties (three from each of the –NH$_3^+$ and –CH$_3$ groups, Figure 1). These six hydrogen bonds are collectively and simultaneously responsible for the titling of the PbBr$_6^{4-}$ octahedra, the contraction of the C–N bond in MA, the N–H/C–H frequency shifts and the stability of the entire static equilibrium geometry of o-MAPbBr$_3$. Other σ-hole (viz. carbon- and pnictogen-bonded interactions) and dispersion dominant Br···Br non-covalent interactions have to be present in this polymorph for the fixed alignment of the organic cation, and for its occupation at the center of the perovskite cage. In addition, we have demonstrated that there cannot be any such indirect-to-direct bandgap transition in o-MAPbBr$_3$ even if we incorporate the effect of SOC, as previously suggested [22].

**Tetragonal MAPbBr$_3$**

Figure 5 shows the PBE relaxed geometry, lattice constants and unit-cell volume of tetragonal CH$_3$NH$_3$PbBr$_3$. The calculated lattice constants are in good agreement with the experimental values, $a = b = 8.322$ Å and $c = 11.833$ Å [111].

The unit cell volume increases from 853.67 to 869.41 Å$^3$ as one moves from the orthorhombic to the tetragonal polymorph. This is a consequence of temperature; as the temperature of the system increases above 162 K there is volume expansion of the perovskite cage consequent on a decrease in the extent of tilting of the PbBr$_6^{4-}$ octahedra in t-MAPbBr$_3$ compared to o-MAPbBr$_3$.

Chen *et al.* [112] have suggested, using group theory, that the MA cation in tetragonal environment of CH$_3$NH$_3$PbI$_3$ has 12 crystallographically equivalent sites for hydrogen atoms, and the hydrogen atoms have two kinds of configurations; hence there can be a total degree of 24 rotational degrees of freedom for the cation inside the cage interior. Weller *et al.* have observed that for tetragonal (*I4/mcm*) CH$_3$NH$_3$PbI$_3$ the MA cation adopts four possible orientations (along [110] and equivalent directions) in the unit cell pointing closely towards the center of the distorted perovskite cube face [19].

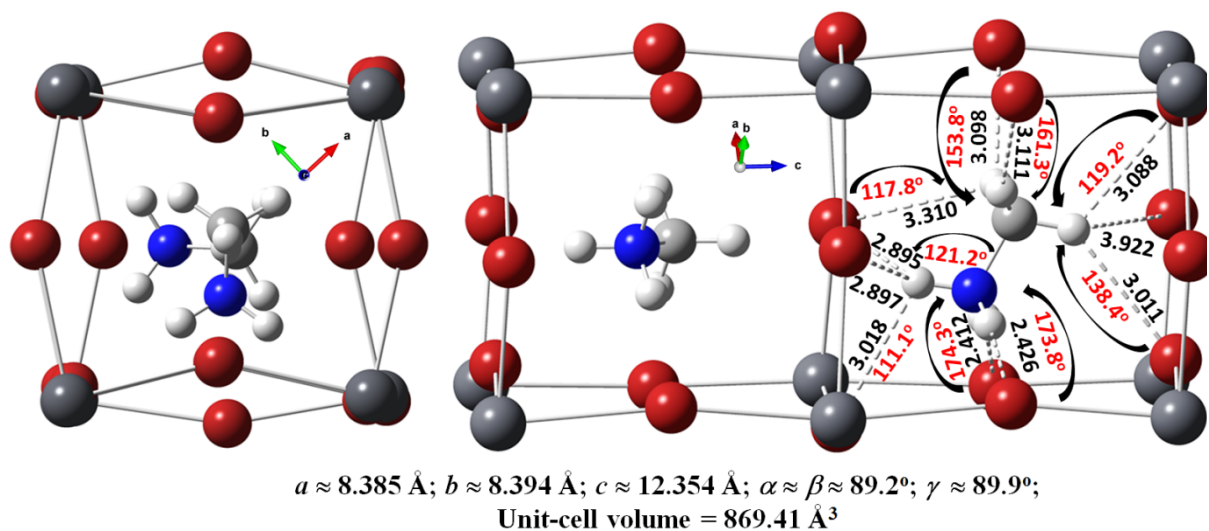

$a \approx 8.385$ Å; $b \approx 8.394$ Å; $c \approx 12.354$ Å; $\alpha \approx \beta \approx 89.2°$; $\gamma \approx 89.9°$;
Unit-cell volume = 869.41 Å$^3$

**Figure 5.** Two views of the ball and stick model of PBE/PAW relaxed geometry of tetragonal MAPbBr$_3$, showing some selected intermolecular distances and bond angles of approach (in Å and deg, respectively). Optimized lattice constants and volume are shown. The energy-minimized unit cell is illustrated in Figure S5.

In the CH$_3$NH$_3$PbBr$_3$ geometry illustrated in Figure 5, the cation is oriented along the distorted [110] direction, an orientation which is similar to that reported experimentally by Swainson *et al.* [15]. As the volume of the perovskite cage in t-CH$_3$NH$_3$PbBr$_3$ increases, the –NH$_3^+$ fragment of the organic cation is slightly pulled toward the wall bromides of the cage to maximize the non-covalent attraction between them. This means the organic cation no longer resides at the center of the perovskite cage; this off-center shift accords with experimental observations.[19-21] As a result, the intermolecular contact distances between the H atoms of –NH$_3^+$ in MA and the Br atoms of the inorganic cage were found to be as short as 2.412, 2.426, 2.895 and 3.018 Å (Figure 5). On the other hand, the non-covalent interaction distances between the H atoms on –CH$_3$ in MA and the coordinated Br atoms lie in a narrow range between 3.011 and 3.310 Å in t-MAPbBr$_3$.

These results indicate that on average the Br⋯H(–N) contacts are stronger than the Br⋯H(–C) contacts in t-MAPbBr$_3$, which is justified since the H atoms on –NH$_3^+$ are significantly more electrophilic than those on the –CH$_3$.

It is worth emphasizing that the "less than the sum of the van der Waals radii" concept is not always robust for identifying intermolecular contacts that are larger than (or very close to) the sum of the van der Waals radii of the putatively interacting atoms (3.06 Å for Br and H atoms) and could produce misleading results [23, 42-44, 46, 96-100, 113]. Use of this criterion alone has failed to identify weakly bound and van der Waals interactions in chemical systems. For instance, past and recent studies show that long (> 3 Å) but directional interactions, which do not obey the aforesaid distance criterion, can be significant in the design of molecular and ordered supramolecular materials [23, 42-44, 46, 96-100, 113], even though these interactions are readily distorted by crystal forces.[44] Han *et al.* have shown that weak interactions, including van der Waals forces, electrostatic repulsions, and halogen bonds, affect the overall structure of the chemical systems they studied [113]. In our perspective, in such cases a combined study of the QTAIM and RDG-NCI topologies of the charge density can be exceptionally informative for the characterization of weak interactions [23, 97-100].

The QTAIM molecular graph for t-MAPbBr$_3$ is displayed in Figure 6; all the intermolecular contact distances shown in Figure 5 are identified as possible hydrogen bonds. However, it suggests each methyl hydrogen is involved either in a bifurcated or in a trifurcated hydrogen bonding topology, forming the Br⋯H(–C) hydrogen bonds. These are confirmed by the presence of bcps and bps between the bonded atomic basins, and smaller values of $\rho_b$ at the bcps. For instance, the $\rho_b$ values at the Br⋯H(–C) bcp's lie between 0.0010 and 0.0072 au, and that this range is somehow consistent with the 0.002 < $\rho_b$ < 0.034 au range recommended to confirm O⋯H(–C) hydrogen bonding on the basis of the charge density at bcps [66].

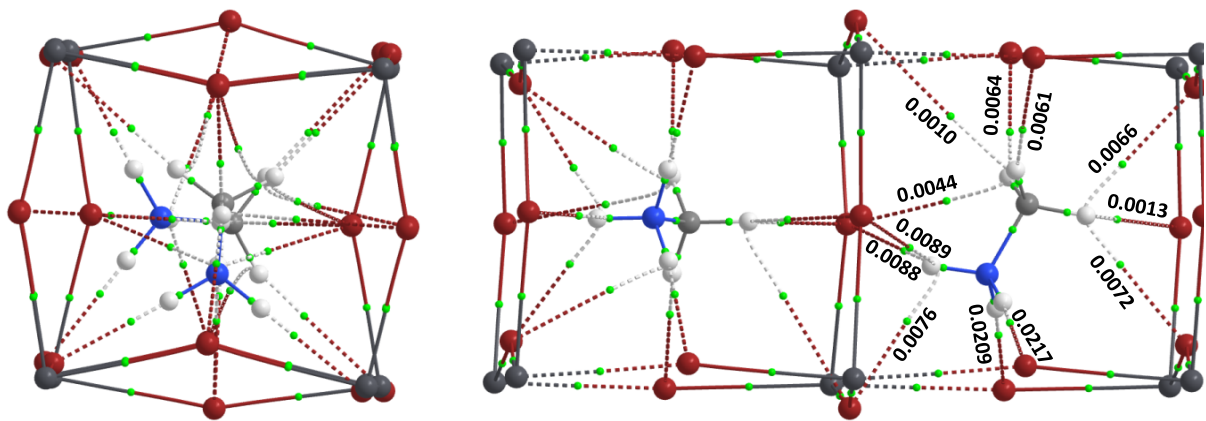

**Figure 6.** Two different views of the PBE level QTAIM molecular graph of the relaxed supercell geometry of t-MAPbBr$_3$, illustrating bcp's (tiny spheres in green) and bond paths (solid and dotted lines in atom color) between various atomic basins interacting with each other. Charge density values at some selected bcp's are shown. Atoms are shown as large and medium spheres. Lead: dark-grey, bromine: dark-red, nitrogen: blue, carbon: grey, and hydrogen: white grey.

The $\nabla^2\rho_b$ values at the Br···H(–C) bcp's are all positive, with values lying between +0.0039 and +0.0205 au, within the range, $0.024 < \nabla^2\rho_b < 0.139$ au, recommended for O···H(–C) hydrogen bonding. The total energy density at these bcp's lies between +0.0003 and +0.0008 au. These results confirm that the Br···H(–C) interactions in t-MAPbBr$_3$ are genuine and closed-shell type [59-65].

Similarly, the –NH$_3^+$ group of MA in t-MAPbBr$_3$ forms two nearly equivalent Br···H(–N) contacts, one with $\rho_b$ = 0.0217 au and the other with $\rho_b$ = 0.0209 au, indicative of an appreciable amount of covalent character in the interaction. This is in agreement with the sign and magnitude of the $\nabla^2\rho_b$ and $H_b$ values calculated at the Br···H(–N) bcp's ($\nabla^2\rho_b$ = +0.0379 and +0.0371 au; $H_b$ = –0.0011 and –0.0010 au). The two signatures are diagnostic of the closed- and opened-shell natures [35-39, 53, 59-65] of the two Br···H(–N) hydrogen bonds in t-MAPbBr$_3$. The remaining H atom on –NH$_3^+$ is found to be involved in a trifurcated hydrogen bonding topology. The strengths of these three Br···H(–N) interactions are comparable ($\rho_b$ = 0.0076, 0.0088 and 0.0089 au; see Figure 5 for intermolecular bond distances associated with these contacts and Figure 6 for detailed bond bp and bcp topologies involved). The $\nabla^2\rho_b$ for the corresponding Br···H(–N) contacts are ca. +0.0236, +0.0245 and +0.0247 au, respectively and the $H_b$ values are +0.0008, +0.0006 and +0.0005 au, respectively. These results show that the Br···H(–N) trifurcated hydrogen bonding interactions formed by a hydrogen of the –NH$_3$ group are predominantly of the closed-shell type [59-65].

The PBE bandgap for t-CH$_3$NH$_3$PbBr$_3$ is calculated to be 2.01 eV, which decreases if SOC is included. The details of the electronic band structure of the system, without and with SOC, are shown in Figure 4b. This result is consistent with the fact that methylammonium lead halide perovskites are considered as direct bandgap semiconductors [114].

In summary, it is evident that the intermolecular bonding topologies in t-MAPbBr$_3$ are not very simple compared to those demonstrated above for o-MAPbBr$_3$ and lacks the rich variety of non-covalent interactions of the latter, which could possibly be due to the different orientation of the organic cation within the cage interior and expansion of the cage volume. The stabilization of the t-MAPbBr$_3$ structure is shown to be driven by the joint effects of both the Br⋯H(–N) and Br⋯H(–C) hydrogen bonding interactions, a conclusion that is contrary to that reported by others [22].

**Pseudo-cubic MAPbBr$_3$**

Figure 7 depicts two energy-minimized configurations of c-MAPbBr$_3$. Their lattice constants and cell volumes are also listed for comparison. The primary difference between the two configurations, a) and b), is that in a) the organic cation is along the [100] direction, whereas in b) it is along the [111] direction. The geometry of the former configuration is about 0.013 eV more stable than the latter; this preference in the morphological stability is in line with the DFT result of Yin *et al.* [22] Since both these configurations have been found experimentally for c-MAPbI$_3$ [19-21, 107, 115, 116], it can be expected that both of them are also experimentally reachable for c-MAPbBr$_3$.

The calculated lattice constants for geometry b) of c-MAPbBr$_3$ are $a = b = c = 6.076$ Å and $\alpha = \beta = \gamma = 89.3°$; hence the geometry of this conformation is nearly but not perfectly cubic. On the other hand, the geometry a) of MAPbBr$_3$ is more distorted, which is evident in the lattice constants, for which, $a \neq b \neq c$ and $\alpha = \gamma \neq \beta$ (Figure 7). While some refer to this geometry as cubic/pseudo-cubic,[105, 117] others refer to it as orthorhombic [118]. The experimental lattice constants for pseudo-cubic MAPbBr$_3$ are $a = b = c = 5.901$ Å [101], with which our results, depicted in Figure 7b, are in reasonable agreement.

Similarly to what was found for the orthorhombic and tetragonal polymorphic geometries, our calculation gave two dominant types of intermolecular contacts between MA and PbBr$_6^{4-}$ for both the pseudo-cubic geometries of MAPbBr$_3$. One type is Br⋯H(–N) and the other type is Br⋯H(–C), with the former are more directional than the latter in both the geometries (Figure 7). The directionality feature is opposite to that observed for the corresponding hydrogen bonded interactions identified in o-MAPbBr$_3$ (Figure 1). This emphasizes the temperature dependence of hydrogen bond directionality.

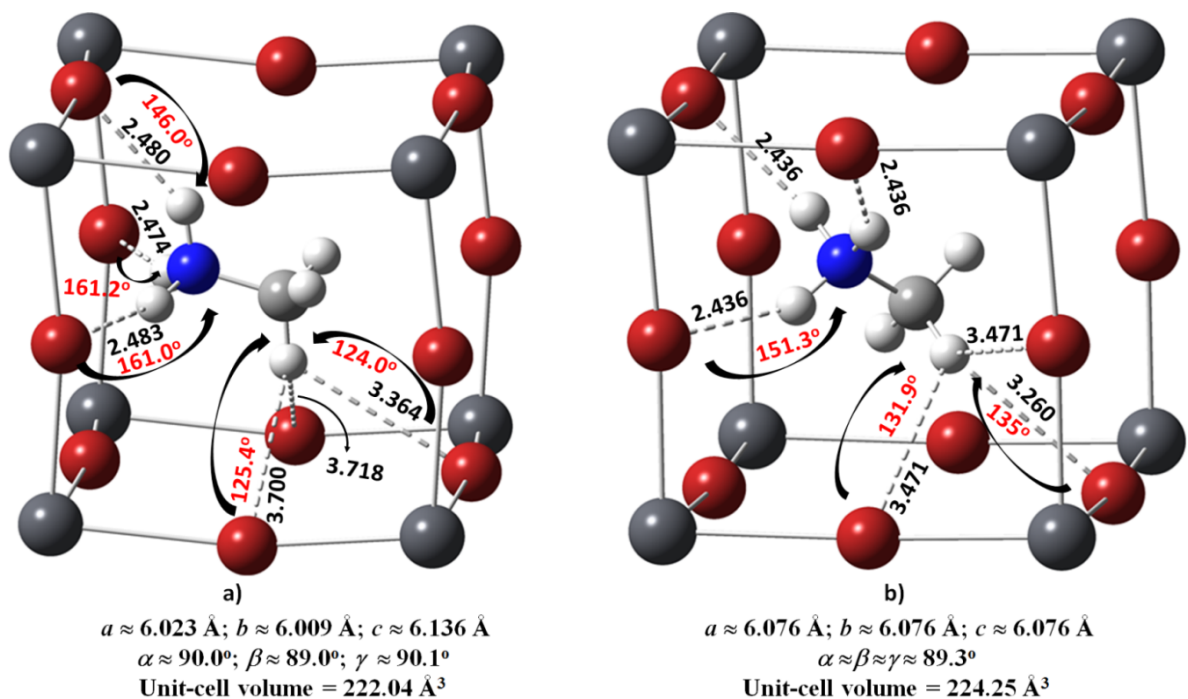

**Figure 7.** Ball and stick models of PBE relaxed 2×2×2 supercell geometries of pseudo-cubic MAPbBr$_3$, showing the orientation of MA along the a) [100] and b) [111] directions, respectively. Intermolecular distances of separation (in Å) and angels of approach (in deg) for putative hydrogen bond formation are shown. (A similar intermolecular bonding topology as shown for a single C–H bond is also expected for the other two H atoms of the –CH$_3$ group). Optimized lattice constants and unit-cell volumes are shown. Orientational details of MA inside the cuboctahedra are also shown in Figs. S6 and S7.

For c-MAPbBr$_3$ with the MA cation aligned along the [111] direction (Figure 7b), each proton on –NH$_3^+$ in MA produces a Br⋯H(–N) contact. There are three such contacts, each with an intermolecular bond distance of 2.436 Å. Each of these distances represents a genuinely strong hydrogen bond since each is much smaller than the sum of the van der Waals radii of 3.06 Å for the Br and H atoms (Br = 1.86 and H = 1.20 Å[82]). This is also supported by the results of the QTAIM analysis (Figure 8), which showed that each Br⋯H(–N) is accompanied with a (3,–1) bcp and a bond path. The $\rho_b$, $\nabla^2\rho_b$, $V_b$, $G_b$ and $H_b$ values at each Br⋯H(–N) bcp are 0.0207, +0.0396, –0.0117, +0.0108 and –0.0009 au, respectively. These signatures ($\nabla^2\rho_b$ > 0 and $H_b$ > 0) at the bcps suggest that each Br⋯H(–N) have appreciable covalency [72,73].

In Figure 7b, the orientation of each methyl H is such that each C–H bond is aligned along the [111] direction, analogous to how the C–N bond of the MA cation is aligned inside the cuboctahedron. Because of this arrangement, two of the Br⋯H(–C) contact distances formed between each of these H atoms and the two nearest Br atoms are equivalent, while the remaining one is somewhat smaller (values: 3.471 *vs.* 3.260 Å). Although these intermolecular distances are significantly longer than the Br⋯H(–N) hydrogen bond distances, QTAIM characterizes each as an intermolecular non-covalent interaction, Figure 8b, since each features a bcp and a bp between H and C atomic basins. This sort of bonding interaction of each H atom to several Br atoms of the perovskite cage may be represented as a bonded cone of density. The $\rho_b$ values for the two types of Br⋯H(–C) hydrogen bonded interactions are 0.0031 and 0.0046 au, respectively. The $\nabla^2\rho_b$, $V_b$, $G_b$ and $H_b$ for each of the two equivalent Br⋯H(–C) interactions are ca. +0.0101, –0.0013, +0.0019 and +0.0006 au, respectively, whereas those for the remaining interaction are +0.0136, –0.0022, +0.0028, and +0.0006 au, respectively. Since the signs of $\nabla^2\rho_b$ and $H_b$ are both positive for these interactions, they are of the closed-shell type [35, 36, 38-40, 59-66, 119, 120]. For both the [100] and [111] orientations of the C–N bond in c-MAPbBr$_3$, the charge density topologies associated with the bonding modes of each H atom of the –CH$_3$ group are very similar. This is evident in Figure 8. For the molecular graph illustrated in Figure 8a, some of the individual Br⋯H(–C) contacts between the interacting H and Br domains are found to be strained. Some of them include bond paths that do not link the H atoms of the –CH$_3$ group to the coordinated Br atoms, rather connecting the vicinity their bcps to the highly electropositive carbon atom of the MA species, thereby producing an outwardly curved bond path between the outer electrostatic surface of the carbon and Br atoms. These are characteristics of the carbon or tetrel bonds [92-94]. The consequence of these strained, bent bonds and their behavior in organic molecules has been described elsewhere [38, 53, 121, 122]. These are readily broken either at the instance of zero-point vibrational motion, or by increasing the temperature of the system as this increases the thermal motion of the organic cation inside the perovskite cage.

The Br⋯H(–N) hydrogen bonds in the pseudo-cubic geometry of Figure 7a, on the other hand, are found to be significantly stronger. There are three primary Br⋯H(–N) hydrogen bonds formed by the three H atoms of the –NH$_3^+$ group and the adjacent bromides of the cage, with each having a bcp charge density of 0.0190 au (Figure 8a). The H atom facing the viewer in Figure 8a is involved in a trifurcated bonding topology, with $\rho_b$ values of 0.0026, 0.0026 and 0.0191 au. In addition, the nitrogen of –NH$_3^+$ group and the nearest Br atoms of the cage are engaged attractively with each other to form the Br⋯H(–N) contacts. These are the characteristics of pnictogen bonding [12, 13]. Two of these latter contacts are equivalent, each with $\rho_b$ = 0.0053 au, and thus contributing strength to the overall geometry of pseudo-cubic MAPbBr$_3$ [100].

In summary, we have shown that both the Br···H(–N) and Br···H(–C) hydrogen bonding interactions are ubiquitous in c-MAPbBr$_3$ (regardless of the two polymorphs examined), meaning that both these interactions collectively are responsible for the c-MAPbBr$_3$ stabilities and their functional properties. In the polymorph with the distorted [100] orientation of the organic cation, carbon- and pnictogen bonds are evident; in fact, these directional interactions appear due to the specific orientation of organic cation, as found for o-M APbBr$_3$. This finding is in sharp contrast to those of others [22], where the presence and importance of the Br···H(–C) and other non-covalent interactions in c-MAPbBr$_3$ were completely overlooked.

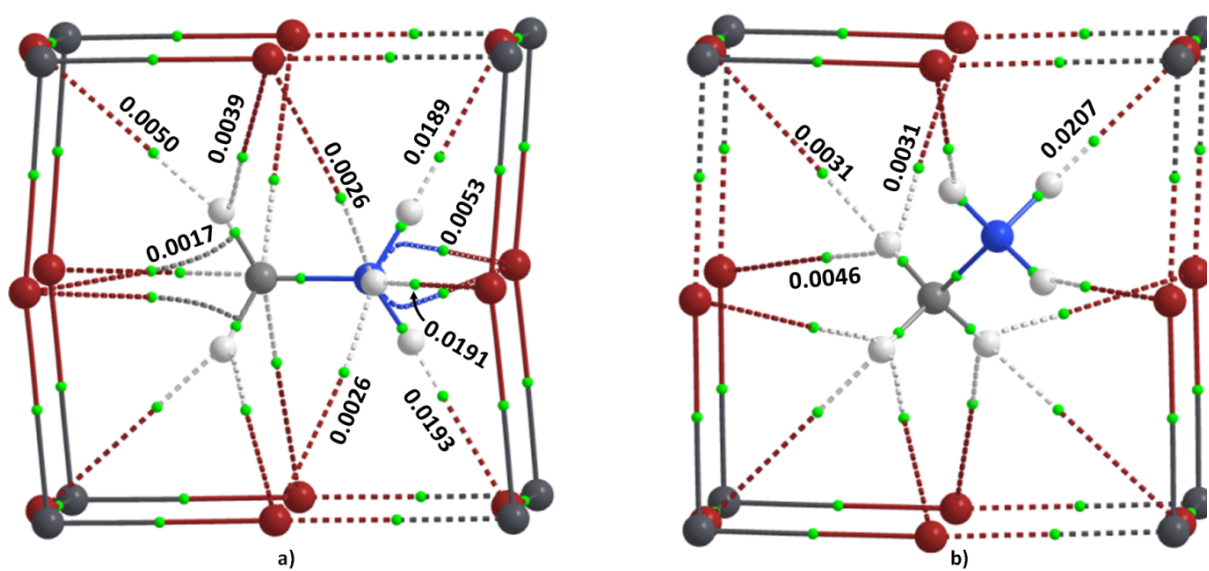

**Figure 8.** PBE level QTAIM molecular graphs of the supercell relaxed geometries of MAPbBr$_3$, with the MA cation aligned along the a) [100] and b) [111] directions, illustrating the bond critical points (tiny spheres in green) and bond paths (solid and dotted lines in atom color). Charge density values at some selected bond critical points are shown. Atoms are shown as large and medium spheres. Lead: dark-grey, bromine: dark-red, nitrogen: blue, carbon: grey, and hydrogen: white grey.

**A Reduced Density Gradient (RDG) description of non-covalent interactions**

In addition to QTAIM, there are several other current state-of-the-art computational tools available to explore non-covalent interactions in chemical systems. The RDG-NCI approach is a robust method [41]. It relies on an examination of the nature of the reduced density gradient *s* around the bond critical point regions as identified using QTAIM, and is described by the relationship given in Eqn 1.

As one moves far away from the nuclei of a system, $s$ will have larger values in regions where $\rho(r)$ decays to zero exponentially, and where the $\rho(r)^{4/3}$ term tends to zero faster than the $|\nabla\rho(r)|$ term. At the bond critical point between two bonded atomic basins in compounds, we have $|\nabla\rho(r)| = 0$, and hence $s = 0$ (the lower bound of RDG). Unequivocally, the bcp is the point where one finds no difference between QTAIM and RDG.

Figure 9 shows the isosurface plot for o-MAPbBr$_3$, t-MAPbBr$_3$, and c-MAPbBr$_3$ [100], obtained with RDG-NCI analysis. Both the QTAIM and van der Waals radii distance-based characterizations given above for the Br⋯H(N), Br⋯H(C), Br⋯C(N), Br⋯N(C), and Br⋯Br interactions are evident in these RDG-NCI plots. For instance, the Pb—Br coordinate bonds, which are QTAIM characterized to be the strongest interactions within the inorganic perovskite framework, are described by disc-like circular RDG volumes in a blue-cyan color – indicating that these have both covalent and ionic character. By contrast, the intermolecular interactions between the organic and inorganic moieties are described by RDG domains of different flavors. The Br⋯H(C) interactions are described by circular RDG domains in green, whereas the Br⋯H(N) interactions are described by circular RDG domains in blue-green. The size of these domains are determined by the strength of the charge density delocalization in the critical bonding regions.

The Br⋯C(N) and Br⋯N(C) carbon and pnictogen bonds, identified for the first time in this study, are found in o-MAPbBr$_3$ and c-MAPbBr$_3$ [100] by QTAIM. They are also evident in the RDG-NCI plot of t-MAPbBr$_3$, but are very weak. These appear as irregular flat RDG domains between N and Br, and between C and Br atomic domains in the RDG-NCI plots illustrated Figure 9 a) and c) for o-MAPbBr$_3$ and c-MAPbBr$_3$ [100], respectively, and are marked as van der Waals in t-MAPbBr$_3$ since such interactions are relatively weak in this geometry. QTAIM did not detect the latter because it sometimes underestimates van der Waals (and very weakly bound) interactions because it misses the bond path and critical point topologies in critical bonding regions. RDG-NCI has identified these because it enables exploration of the gradient of charge density in the close vicinity of the bcp. This has been elucidated in several recent studies by the Kjaergaard group [67-71], in which it was shown that the experimentally-determined weakly bound and van der Waals interactions in molecular systems can be successfully identified and characterized by RDG-NCI, which may or may not be identified with QTAIM due to its stringent criteria.

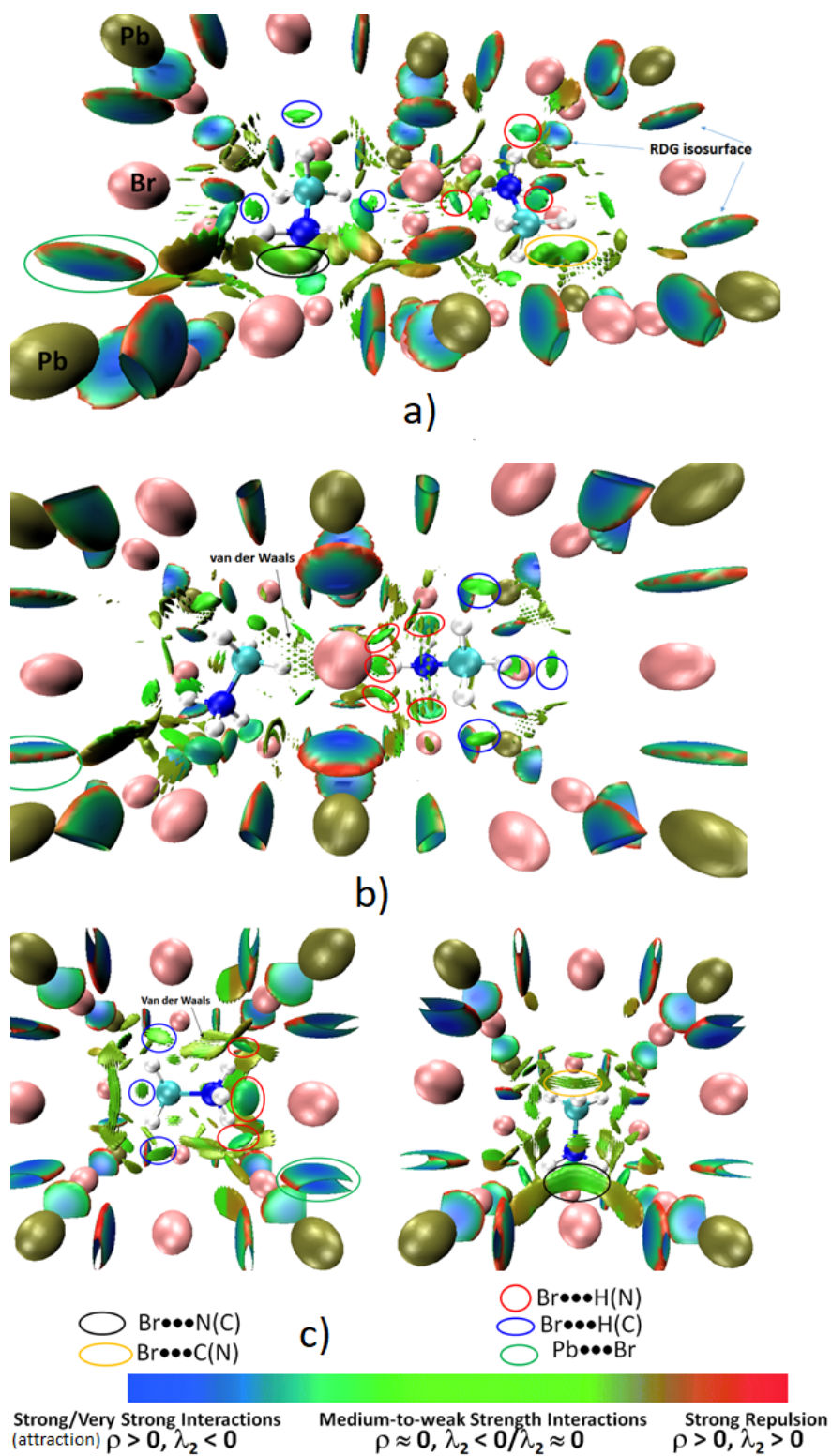

**Figure 9.** PBE level RDG-NCI isosurface plot of a) o-$CH_3NH_3PbBr_3$, b) t-$CH_3NH_3PbBr_3$, and c) c-$CH_3NH_3PbBr_3$ [100]. These are obtained by calculations on their bulk/supercell geometries (RDG isovalue = 0.5 a.u.). For the isosurface plot in c), two different views are shown for clarity. Isosurfaces between various atomic domains are illustrated in different colors. The regions on the color bar represent the strength of the different types of interactions, which are determined by the strength of $\rho$ and sign of the second eigenvalue $\lambda_2$ of the Hessian of charge density matrix. Several most important isosurfaces are circled to show the presence of some specific interactions (see text).

**General discussion on hydrogen bonding**

It is well known that intermolecular bond distance is one of the most important attributes for understanding any kind of non-covalent bonding interaction in chemical systems. It is responsible not only for local and global stabilities but also for giving an overall shape to any complex chemical system (a DNA strand, for example [14, 123, 124]). Although many definitions and characterizing features have been noted in the large body of work that explores hydrogen bonding in chemical systems (for example, Refs. [8, 23, 42, 44, 125-133]) Parthasarathi *et al.* [58] have used QTAIM charge density results and Desiraju [42, 44, 133] has used (tentative) energy and distance based characteristics to recognize that hydrogen bonds have no strict boundaries, thus setting such definitive energetic or distance boundaries for non-covalent interactions may be unproductive.

It has been recommended that for strong hydrogen bonds the intermolecular bond distances should vary between 1.2 and 1.5 Å; for moderately strong hydrogen bonds such distances should vary between 1.5 and 2.2 Å; and for weak bonds such distances can go beyond 2.2 Å [8, 23, 42, 44, 125-133]. However, we have just recently demonstrated that the Y⋯H (Y = F, Cl, Br, I) hydrogen bonds in $BMY_3$ organic-inorganic hybrid perovskites do not obey any such bond distance boundary condition [23] as recommended in general for moderate-to-strong hydrogen bonds [8, 23, 42, 44, 125-133]. In particular, we have shown that within the bimolecular approximation the Y⋯H(–N) and Y⋯H(–C) hydrogen bonds in $BMY_3$ perovskite building blocks can be unusually strong, with binding energies much larger than the covalent limit (ca. –40 kcal mol$^{-1}$). The Y⋯H intermolecular bond distances for these perovskites were computed to be in the 2.0–2.7 Å range. While the same range has been proposed in general for weakly bound interactions [8, 23, 42, 44, 125-133], we have suggested that the Y⋯H contacts in $BMY_3$ perovskites (with the same range of intermolecular distances of separation) be referred to as *ultra-strong* hydrogen bonds [23, 24].

The characteristics and classification of hydrogen bonds documented in the literature emerge from the A⋯H–D motif, where the proton donor fragments D are the π-systems, O, N, S, F, Cl, *etc.*, and the proton acceptor fragments A are mainly, but not limited to, O, N, S and F. Desiraju [44, 133] and others [129, 130, 134] have comprehensively discussed the importance of O⋯H(–C) hydrogen bonds, containing the low polarity C–H bond, in biology and crystal design. These are directional interactions, but the intermolecular angles for approach of the electrophile have been normally reported in the range between 90 and 180°.

The intermolecular bond distances for such hydrogen bonds are usually smaller than 3.2 Å [130]. This value has been frequently used as the cutoff distance for hydrogen bond identification, even though it is much larger than the sum of the van der Waals radii of the O and H atoms (2.7 Å). To give an example, the crystal structure of the [Cr(CO)$_3$[η$^6$-[7-*exo*-(C≡CH)C$_7$H$_7$]]] compound contains a very long O···H(–C) contact from the ethynyl group to a carbonyl ligand with a H···O separation of 2.92 Å; the weakly hydrogen bonding character of this contact is inferred from the Raman and the IR absorption spectra, showing the long range nature of the aforementioned O···H(–C) hydrogen bond [43]. There are several other examples documented in the literature that fall into the same category. For example, the O···I(−C) halogen bonding distance reported in some complex systems to be as large as 4.2 Å and O···I−C angle of approach to be 139°, even though the O···I distance is much larger than the sum 2.54 Å of the van der Waals radii of the O (1.50 Å) and I (2.04 Å) atoms forming the contact [134]. Not only for this bond, but for weakly bound interactions in general the distance cut-off has been recommended to be in the 2.2 – 3.2 Å range [8, 23, 42, 44, 125-133]. Thus, given that the O atom is much less dispersive and polarizable than heavy atoms such as the halogens Br and I, given that the van der Waals radii of these halogens are larger than the O atom (1.50 Å), and given that the recommended intermolecular bond distances for O···H(–C) hydrogen bonds can be as large as 3.2 Å, it is evident that the Br···H(–C) intermolecular bond distances close to the value of 3.2 Å (or even longer) as found in the various polymorphs of MAPbBr$_3$ are genuine hydrogen bonds.

It seems clear that to prescribe that the intermolecular distance must lie between 1.2 and 1.5 Å in any molecular complex featuring strongly bound Br···H and I···H contacts (as recommended in general for strong and very strong hydrogen bonds[8, 23, 42, 44, 125-133] is untenable. If we take into account the van der Waals radii of 1.20, 1.86, and 2.04 Å for the H, Br and I atoms, respectively [82], then for very strong Br···H or I···H hydrogen bonds the respective intermolecular bonded distance would be close to the van der Waals radius of either the Br or the I atom (1.86 Å and 2.04 Å); these values cannot lie in the 1.2–1.5 Å range. Similarly, for weaker Br···H and I···H hydrogen bonds, the corresponding intermolecular distances will be comparable to, or even larger than, the sum of the van der Waals radii of the X (X = Br and I) and H atoms, e.g., 3.06 and 3.24 Å for Br···H and I···H, respectively. Since the Br and I atoms are heavier compared to the main group elements such as N, O and F, *etc.*, it is expected that the hydrogen bond energies formed by these halogens would involve contributions from dispersion and polarization interactions that will play a significant role in the effective stabilization of any weakly bound interaction. Moreover, if we consider the experimental uncertainty in the proposed van der Waals radii of atoms to be ± 0.2 Å, then there must be some flexibility in the interatomic bond distances speculated above between the H and X atoms. Desiraju has demonstrated on several occasions that O···H(–C) intermolecular bond distances ranging between 3.0 and 4.0 Å for D···A are not only characterizeable as hydrogen bonds, but also as important motifs for the geometrical stabilization of chemical compounds [42]. The same argument is applicable to the Br···H(–C) hydrogen bonds in $CH_3NH_3PbBr_3$ as well. If we consider the van der Waals distance between the Br and C atoms to be 3.63 Å, the sum of the van der Waals radii of the Br and C atoms (1.86 and 1.77 Å, respectively), then this value is typically less than the D···A distance cutoffs of 3.70 Å, which has been frequently used as a metric for the identification of C···H hydrogen bonds [134].

Furthermore, halogen centered non-covalent interactions are both electrostatically and dispersion dominant [135], especially when these involve I and Br atoms. Because these atoms are readily polarizable (Cl < Br < I), it is not always easy to predict under what circumstance halogen atoms in molecules attract the negative and positive site to form a non-covalent interaction [136, 137]. For instance, it has been shown on some occasions that halogen, such as fluorine, in molecules can be entirely negative [96]. With such a potential negative profile, it has shown its ability to form non-covalent interactions with another negative site(s) to form ordered crystals [96]. This is facilitated by the underexplored London dispersion that has an $r^{-6}$ dependence [96-100]. Analogous arguments have provided for non-covalent interactions formed by other halogen derivatives [100]. Taking all these anomalous features into account, it is clear that the range of values recommended for $\rho_b$ and $\nabla^2\rho_b$ (0.002 < $\rho_b$ < 0.034 au; 0.024 < $\nabla^2\rho_b$ < 0.139 au) to identify and characterize O···H(–C) hydrogen bonding in molecular complexes [66] should not be invoked *as the sole criterion* to determine whether all kinds of hydrogen-, σ-hole and dispersion assisted bonding interactions exist in chemical systems that are formed by donors of various types. The sole use of these values without showing the presence of a bcp and a bp (an integral part of the molecular graph and charge distributions) is misleading since these values are counter to the QTAIM requirements deemed necessary to infer chemical bonding. Also, the $\rho_b$ and $\nabla^2\rho_b$ values recommended for O···H(–C) hydrogen bonding obtained with a given functional and basis set are likely to be affected by changing the level of theory and basis set as are not invariant; these values are therefore not robust. We have illustrated one of these features for the systems presented here in which the use of these values for $\rho_b$ and $\nabla^2\rho_b$ has underestimated the presence of a number of important intermolecular interactions, regardless of whether these concern hydrogen bonding or any other non-covalent interaction, which were subsequently identified by the signatures of QTAIM and RDG-NCI based charge density topologies.

## 4. Conclusions

This study has enabled us to discuss the detailed nature of the evolution of various non-covalent interactions between the organic and inorganic frameworks responsible for the polymorphic transformations of MAPbBr$_3$. The conclusions drawn are markedly in contrast to a variety of interpretations and findings that have already been documented in the halide perovskite literature; these were unduly and incorrectly biased to hydrogen bonds formed by the ammonium fragment of the organic cation. In particular, we have shown that in addition to hydrogen bonds formed by the ammonium and methyl groups there are other σ-hole and dispersion-dominant interactions that play an important role in the observed octahedral tilting in o-MAPbBr$_3$, and for the matching between the center-of-mass of the MA cation and the center of the perovskite cage, as well as for the fixed orientation and alignment of the organic cation inside the perovskite cage in o-MAPbBr$_3$. We have demonstrated that there cannot be any indirect-to-direct bandgap transition in o-MAPbBr$_3$, as some have suggested. The nature of the bandgap should always be direct at $\Gamma$, regardless of whether the electronic structure calculations are carried out with or without SOC, and regardless of the nature of the various intermolecular non-covalent interactions involved in the development of o- and t-MAPbBr$_3$. However, for the two c-MAPbBr$_3$ polymorphs, we have indeed shown direct-to-indirect bandgap transitions; consequently, this material has displayed outstanding functional properties for applications in optoelectronics. We have also shown that this bandgap transition is the repercussion of SOC, but is not unexpected when heavy atoms such as Pb are involved. Our finding has also suggested that the direct-to-indirect bandgap transition should not just be attributable to the Br···H(–C) hydrogen bonding interactions, as discussed in a few other studies. If such an argument is valid, then the same argument should also be applicable to the Br···H(–N) hydrogen bonding interactions, as well as to o-MAPbBr$_3$ since the sole existence of this polymorph is driven by the two types of hydrogen bonded interactions, as well as by other non-covalent interactions, that are highly competitive. While the dynamic aspect is not taken into account as it is unnecessary in the present context, it is indeed not very hard to infer what relevance these findings have for perovskite materials. A background study might be necessary to infer the similarities and differences, especially when one focusses discussion on chemical bonding in perovskite materials. Otherwise it might lead to the view that the results emanating from the models adopted in this study would contravene what one knows about the existing non-covalent physical chemistry of perovskites and the dynamics involved.

Some have doubted whether the strength of "hydrogen bonds" between a hydrogen involved in a low-polarity bond such as C–H bond, and a large halogen atom such as Br can be evaluated. We have provided various arguments in support of this interaction, and have discussed possible occasions that would assist in characterizing this interaction in large-scale systems. We have shown that the formation of both the Br⋯H(–C) and Br⋯H(–N) hydrogen bonds are ubiquitous in the tetragonal and pseudo-cubic polymorphs, as is the case for the orthorhombic polymorph of MAPbBr$_3$. That is, the Br⋯H(–C) hydrogen bonds can be understood as an inherent characteristic of the MAPbBr$_3$ system, which appear regardless of the nature of the size, volume and lattice expansion or contraction associated with the three polymorphs of the MAPbBr$_3$ system. We note that at room temperature many different studies coming from neutron scattering, Kerr effect, or time-resolve IR, show that the MA cation is weakly bound and experiences fast rotation inside the perovskite cage, but how, and in which way, does it remain weakly bound remains unclear. Some suggest the real characterization of the MA cation is undetermined, since it is difficult to detect small atoms in the presence of several heavier ones. Nevertheless, in addition to the aforesaid interactions, the presence of the carbon- and pnictogen bonding interactions are shown to appear depending on the orientational freedom of the organic cation, including the c-MAPbBr$_3$ [100] polymorph. Although weaker in the room temperature polymorphs, the importance of Br⋯H(–C) interactions should not be overlooked, as is often done in the perovskite literature, as they assist not only in stabilizing the overall geometry of each of the three phases of the MAPbBr$_3$ perovskite system as a driving force for the polymorphic transformation, but also in the eventual evolution of the optoelectronic properties required for the development of perovskite materials for photovoltaic applications. Without these hydrogen bonds and/or the other non-covalent interactions revealed, no matter how strong and weak they are, and how static and dynamic they are, halide-based perovskites would have no real physical existence.

**Acknowledgements**

AV, PRV and KY thank Institute of Molecular Science, Okazaki, Japan for supercomputing facilities received for all calculations, and thank CREST project for generous funding (Grant No. JPMJCR12C4). HMM thanks the National Research Foundation, Pretoria, South Africa, and the University of the Witwatersrand for funding.

## Conflict of interest

The authors declare no conflict of interest.